\documentstyle[aps,prl]{revtex}

\begin{document}

\title{Averaging Einstein's Equations: The Linearized Case} 

\author{William R. Stoeger$^{\star}$, Amina Helmi$^{\dag}$, 
and Diego F. Torres$^{\ddag}$ }

\address{$^{\star}$ Vatican Observatory Research Group, Steward Observatory,
Tucson, AZ 85721}

\address{$^{\dag}$ Sterrewacht Leiden, University of Leiden, Niels Bohrweg 2,
2333 CA Leiden, The Netherlands}

\address{$^{\ddag}$ Departamento de F\'{\i}sica,  Universidad Nacional de La
Plata, C.C. 67, 1900 La Plata, Argentina}

\maketitle
\begin{abstract}
We introduce a simple and 
straight-forward averaging procedure, which is a
generalization of one which is commonly used in electrodynamics, and show
that it possesses all the characteristics we require for linearized averaging
in general relativity and cosmology -- for weak-field and perturbed FLRW
situations. In particular we demonstrate that it yields quantities which
are approximately tensorial in these situations, and that its application to
an exact FLRW metric yields another FLRW metric, to first-order in integrals
over the local coordinates. Finally, we indicate some important limits
of any linearized averaging procedure with respect to cosmological 
perturbations which are the result of averages over large amplitude small
and intermediate scale inhomogeneities, and show our averaging procedure
can be approximately implemented by that of Zotov and Stoeger in these cases.
\end{abstract}

\section{Introduction}

It is usual in cosmology to consider
the standard Universe as spatially homogeneous and isotropic
on the largest scales. In fact, there is very good observational support for
doing so. However, we also know that inhomogeneities
exist at almost all scales -- the smaller the scale the larger the 
inhomogeneity.
All these inhomogeneities are apparently insignificant in cosmology, as long
as we are not interested in modelling structure formation. It is therefore
usual
and considered acceptable by the vast majority of researchers
to ignore them in investigating the dynamics and geometry of the
Universe as a whole. Nevertheless, Ellis \cite{Ellis} has given compelling
reasons why conceptually the large-scale cosmological metric should really be
an average over very large regions of space-time. According to this view,
the metric $g_{\mu \nu}$ and the Einstein field equations
\begin{equation}
G_{\mu \nu} \equiv R_{\mu \nu} - \frac{1}{2}g_{\mu \nu} R = 8 \pi T_{\mu \nu},
\end{equation}
where $R_{\mu \nu}$ is the Ricci tensor and $R$ is the Ricci scalar, both
depending on second-order derivatives of $g_{\mu \nu}$, and $T_{\mu \nu}$ is
the stress-energy tensor, must be averaged over
small and intermediate local
inhomogeneities on larger and larger scales to
obtain the average cosmological metric and the averaged 
dynamical equations. \\

However, 
averaging and operating with the Einstein differential operator on a metric
do not commute, because of the nonlinearity of the operator. Thus, the
solution to the averaged Einstein equations will {\it not} be the averaged
metric. The averaged metric will, therefore, 
obey equations
different from the averaged Einstein equations. 
In general, the smoothing-out operation will introduce
extra tensor terms in the field equations, which may affect
the dynamics and the energy conditions in the averaged universe.
Furthermore and just as importantly, it is a very complicated and unresolved
issue to define an adequate averaging
scheme with  the necessary properties -- including the uniqueness of the 
averaged objects and their at least approximate tensorial character. 
This difficulty arises,
because,  in general, integrating a tensor field does not yield another tensor
field in a curved space-time. \\

In spite of these difficulties and uncertainties, 
several averaging procedures have been proposed
by Isaacson \cite{Isaac}, Noonan \cite{Noonan1,Noonan2}, 
Zotov and Stoeger \cite{Zotov}, and more recently 
by Boersma \cite{Boersma}.
Their different philosophies and results, together with a comparison with the
averaging scheme we shall introduce here, 
are briefly described and discussed
below. Although far from the object of the present paper, 
it is also worth mentioning the 
ideas developed by Zalaletdinov in constructing his theory
of macroscopic gravity \cite{Zalaletdinov}. 
He has used the concept of duality -- the existence of two types of observers,
microscopic and macroscopic -- for the study of classical physical phenomena.
Relying on this, he constructed a theory based on averaging a curved
space-time itself
and then  determined the geometrical objects (metric, connection and
curvature)which describe the averaged space-time \cite{Zalaletdinov}.
This approach is non-pertubative in nature.\\

In this investigation we define an averaging scheme, 
which is a straight-forward 
generalization of one often used in macroscopic electromagnetic
theory \cite{Jackson}, 
carefully examine its properties, particularly for weak fields
and for perturbations, and show that it is also an improvement of
Noonan's averaging procedure
\cite{Noonan1,Noonan2}. The behavior of our averaging scheme
in the cases in which the linear approximation 
is sufficient is acceptable. In these
situations the noncommutability of averaging and operating on the
metric with the relevant differential operator is replaced by commutability,
yielding simple, almost trivial results. However, these provide an essential
reference for evaluating our averaging procedure and for reaching firm
conclusions concerning the 
proper interpretation of and constraints on cosmological averaging in general,
as well as interesting and important applications. 
Since much
of cosmology involves perturbation treatments, we believe it is crucial to
be clear about
the averaging operator and its results in this context first, in order to be 
able to understand the cases in  which nonlinear effects become important. \\

Next, we  discuss in detail the general properties of our averaging
operator --
in particular its approximately tensorial character in cosmological
coordinates and for general coordinates in the weak field case, and
its apparent lack of uniqueness with respect to local coordinates. \\

Penultimately, we apply our averaging procedure to the important case of
perturbed 
Friedmann-Lema\^{\i}tre-Robertson-Walker (FLRW) 
cosmologies, showing that averaging an exact FLRW space-time yields, to first
order, in the local coordinates, an FLRW space-time, and further that 
averaging a perturbed FLRW cosmology yields, as expected, another perturbed
FLRW cosmology. We then discuss in some detail, from the averaging point
of view, what type of inhomogeneities in an FLRW background may be considered 
as perturbations of an FLRW metric. Our linear formalism is able to handle
deviations for which the local dynamics are not completely decoupled from the
general expansion of the universe. In contrast, if the local inhomogeneities
are not in the linear regime, then a first order perturbative approach cannot
be applied. Nevertheless, our general averaging scheme is still valid, but the
resulting metric is not a simple superposition of FLRW plus perturbations. \\

Finally, we show how our averaging procedure may be approximately implemented
in this nonperturbative case in simple situations. The procedure used by
Zotov and Stoeger \cite{Zotov}, applied throughout a space at each point,
essentially accomplishes in an easily implementable way what our averaging
procedure requires. 
\section{The Averaging Procedure}

\subsection{Definition}

We define an  averaging operator, or simply, an {\it averager}, acting on a 
field $Q$ as 

\begin{equation}
<Q(x)>=\frac{{\displaystyle \int \! Q(x+x')\, \sqrt{-g(x + x')}
\,d\Omega '}}{{\displaystyle \int \! \sqrt{-g(x + x')} \,d\Omega '}}.
\end{equation}
This is a simple extension of the  definition used by Jackson
\cite{Jackson}, for macroscopic electromagnetic fields.
Here $Q$ is any field -- it may be a scalar, a vector, or a tensor. The
coordinate $x$
gives the cosmic, large-scale location of the averaging volume in space-time,
and $x'$ gives the small-scale
location of a
point within the averaging volume $\Omega'$ relative to its location $x $
(its center, if it is a sphere) in the space-time, both expressed in
Minkowski (rectangular) coordinates. Throughout this paper we often write
$x$ and $x'$ for the cosmological and local
coordinates, respectively, and
$x + x'$ for their sum, by which we mean $x^{\mu}$, $x'^{\mu}$ and 
$x^{\mu} + x'^{\mu}$, respectively.
The metric of the space-time, for the observer
who sees the inhomogeneities, is $g_{\mu \nu}(x+x')$ at 
the point $x + x'$. This is why the determinant of the metric $g$ in the
integrands of 
Eq. (2) depends
on $x + x'$. Though each 
integration is only over the local regions of space-time dominated by the 
inhomogeneities, there are innumerable such averaging volumes spread over the
observable universe -- one for each cosmological coordinate $x$ -- and the
metric within each one depends, of course, not just on the local coordinate
$x'$ but also on the cosmic location of the the averaging volume, given by
the cosmological coordinate $x$. Obviously, the result of the averaging will
depend strongly on the length scale over which it is performed. 
In general, that will be larger than, or of the order of, the
characteristic length scale of the inhomogeneities over which we want to
average.  \\

This definition is not invariant under a change of
coordinates. It can only be implemented as such in a coordinate system like
Minkowski's in which the cosmological coordinates and the local coordinates
are ``parallel'', or translationally related. Nevertheless, as we shall see,
our averager has some very nice properties. In particular, it is
almost invariant under transformations of the cosmological coordinates, and
under transformations of the complete coordinate $x + x'$ in the weak field
and perturbation cases. \\

We would like to stress that
the quantities that are averaged are always referred to the same scale.
In the process of averaging, 
the only observer involved is the one at the local `small
scale', which means that we are treating the problem in a self-consistent way.
In other considerations one might have another observer -- a cosmic
observer -- who does not
see the inhomogeneities. In those cases, we would need to compare
the results of the latter with the averaged ones.  \\

There are two motivations for choosing this procedure for averaging. The
first is simply that it implements what we intuitively envision as 
averaging over a given length scale, providing 
an assignment of an average
value of a quantity to every point $x$ throughout the space-time manifold
taking into consideration the possibly different values of the metric
in different cosmological locations, and specifying unambiguously the 
relationship between the large-scale cosmological coordinate system and the
local, small-scale coordinate system, over which the averages are carried out.
One simply takes the averaging volume and shifts it from point to point
throughout the universe, averaging at each 
point according to  Eq. (2). This is
precisely what is done in making the transition from 
microscopic to macroscopic
electromagnetism 
\cite{Jackson}, as we have indicated above. Secondly, as we
can explicitly show (see Section IV), the averaging procedure defined in
Eq. (2) yields quantities which are approximately tensorial in character
with respect to the cosmological coordinates $x$, and with respect to the full
coordinates $x + x'$ for weak gravitational fields. That is, 
if $Q(x + x')$ is
a tensor with respect to
$x + x'$, $<Q(x)>$ is almost a tensor of the same type with respect to $x$, and
with respect to $x + x'$ for weak fields and perturbed space-times --
deviating from a true tensor by only very small quantities. This is
a very attractive property of our averager.  \\

At the same time, there are some other aspects of our averager  which, at first
sight, are not
so attractive and require further explanation. First, as we have already 
mentioned, the
relationship between the cosmological and the local coordinate systems, as 
formally expressed in Eq. (2), can only
be realized in a very retricted class of coordinate systems
 -- those in which they can be related to
one another by simple translation, so that the ``complete'' coordinate of
any point can be expressed simply as the sum of the cosmological coordinates
and the local coordinates at that point.
Most coordinate systems will not fulfil this requirement. 
This may not be a problem, since, within the demands of coordinate covariance,
we are allowed to choose any coordinate system we want. In particular, any 
coordinate system which simplifies the formulation of our problem may be
selected -- as long as the quantities in question are tensors, or nearly 
tensors, thus assuring us that they are equivalent to their forms in more
complicated coordinate systems.
For other choices we could also specify the relation
between the cosmological coordinates and the local 
coordinates, but the relationship between the two would, in general, be much
more complicated -- not specifiable by a simple sum of the two coordinates,
that is by simple translation. \\

This relation between the cosmological coordinates and the local coordinates 
leads another potential difficulty. Although it is clear that 
the average we have defined in Eq. (2) 
is approximately covariant with
respect to the cosmological coordinates 
\cite{Nelson}, as we shall show in Section V, 
strictly speaking a transformation of the
cosmological coordinates $x$ in the integrals of 
Eq. (2) should be accompanied by a transformation of the 
$x'$ coordinates, along with an 
induced change in their functional relationship (from a simple sum to something
more complicated), as indicated above. However, even then the
averaging over the local coordinates would still not be invariant under those
changes. In light of the arguments given by Ellis and
Matravers \cite{EllisMatr} and
Ellis, Matravers and Zalaletdinov \cite{EllMaZal} for using preferred coordinate
systems in general relativity which are appropriate to the physical situation
and simplify the problem, even though they break coordinate invariance,
we maintain that this lack of covariance of the averaging procedure
with respect to the local coordinates should not be considered an essential
problem in most cosmological applications. As we shall show, this is certainly
the case for
weak gravitational fields and for perturbations from an exact cosmological
solution (e.g. FLRW) to the field equations. In these cases
the deviations caused by the lack of coordinate covariance of the integrals
with respect to the local coodinates are small. Even for more general cases,
it is likely that these deviations will be small, as long as we are averaging
over volumes for which the space is almost flat -- for which the length
scale of the averaging volumes is smaller than the radius of curvature of the
universe. Thus, in that sense the most important property of our averaging
procedure is that it is covariant relative to transformations of the
cosmological coordinates. \\

Another way of expressing the above objection is that in carrying out our
averaging procedure we are effectively
adding vectors and tensors at different points, which is not a well-defined,
or even an allowed, operation on vectors and tensors 
if the space-time is curved. We should really 
incorporate bi-vectors in the averaging integrals in order to translate
vectors and tensors to the point designated by the cosmological metric and then
add them, as proposed by Isaacson \cite{Isaac}. The principal reason why we
have not done this is that there is no easy way of defining the needed
bi-vectors without already knowing the cosmological background metric, and
it is precisely this cosmological background metric that we eventually want
to determine
by carrying out the averaging over small scales. We discuss this in more
detail in the next subsection below. (Even though throughout most of this paper
we limit ourselves to cases in which we have a well-defined background -- the
weak-field and perturbed FLRW situations -- we want to be able to use our
averager in more general cases. We shall present one way of doing that
approximately in the last section.) Furthermore, as we have just discussed,
in the case of weak fields
and of almost flat averaging volumes, the errors introduced by neglecting
the bivectors will be small. Zalaletdinov \cite{Zalaletdinov} constructs his
macroscopic theory of gravity with bi-vectors, but these have to be solved
consistently with the other field equations. This is a possible way of
proceeding, but leads to such a complicated theory that solutions in the 
simplest cases have yet to be obtained. \\ 

\subsection{Comparison with other definitions}

There are a number of other definitions of averaging which have been 
suggested. We describe them briefly here, and compare them with ours, indicating
its advantages. \\ 
\begin{itemize}

\item{Noonan's Operator}

The averaging operator given by Noonan \cite{Noonan1,Noonan2} is
\begin{equation}
<Q^\prime>=\frac{{\displaystyle \int \! Q^\prime(x')\, \sqrt{-g(x')} \,d\Omega'}
}{{\displaystyle \int \! \sqrt{-g}\, d\Omega'}},
\end{equation}
where
\begin{itemize}
\item $\sqrt{-g}$ refers to the determinant of a macroscopic metric. 
\item The region of integration is finite and its size is 
bigger than the usual scale of the 
microscopic observer and smaller than that of the macroscopic observer. Using
this fact the denominator may be approximated by $\int \! d\Omega$ in the 
cases in which
the variation of $g$ is not abrupt.
\item The dependence of $<Q^\prime>$ on the 
space-time coordinates comes from using indefinite integrals. This means
that the variable in the averager is now the boundary of the 
integration volume -- which, we believe, is not well-defined for an averager.
\end{itemize}

The three points mentioned above represent the differences from our definition 
of the averaging procedure. One possible weakness of Noonan's approach (for
details, see also \cite{Roque}) is the simultaneous introduction of two scales
in his averages: the metric in his definition is one of
large scale, whereas his averaging quantities are over small scales. 

\item{Isaccson's Operator}

Isaacson's procedure of averaging \cite{Isaac} directly deals with the
problem that, in general, the result of integrating a tensor field
does not give another tensor, because tensors at different points have
transformation properties which depend on their location, as we have already
indicated. Since one can only add tensors at the same
point, the objective is to carry them to a certain common point and to add
them there. To do that, one has to introduce the {\it bi-vector of parallel
displacement} $j^{\beta}_{\alpha}(x,x^\prime)$ \cite{DeWitt,Synge,Grav}. 
This object transforms
as a vector with respect to coordinate transformations at $x$ or at $x^\prime$,
and it has the property that,
given a vector (or tensor, in general) 
$A_{\beta}$ at $x^\prime$, then 
$A_\alpha(x)=j^{\beta}_{\alpha}(x,x^\prime) A_{\beta}(x^\prime)$ 
is the unique 
vector at $x$ that can be obtained by parallel displacement of 
$A_{\beta}$ from $x$ to $x^\prime$ along a geodesic. \footnote{This is
not true if there is more than one geodesic which joins the two
given points $x$ and $x^\prime$, because the transport along each different 
geodesic varies. One may avoid such problems arguing that the two points
are close enough to each other as to permit the existence of only one geodesic
joining them.} \\

Given a tensor $T_{\mu\nu}$ Isaacson's averaging operator is defined as:

\begin{equation}
<T_{\mu\nu}(x)>=\int_{all\,space} 
j^{\alpha^\prime}_{\mu}(x,x^\prime)
j^{\beta^\prime}_{\nu}(x,x^\prime) T_{\alpha^\prime \beta^\prime}(x^\prime)
f(x,x^\prime) d^4x^\prime
\end{equation}
where $f(x,x^\prime)$ is a weighting function which falls smoothly to zero
when $x$ and $x^\prime$ differ by a distance greater than $d$, its integral is
normalized
to one over all space-time.
This definition carries with it
all the properties of the bi-vector of parallel geodesic displacement, 
for example

\begin{equation}
j^{\alpha^\prime}_{\mu}(x,x^\prime)
j^{\beta^\prime}_{\nu}(x,x^\prime) 
g_{\alpha^\prime \beta^\prime}(x^\prime)=
g_{\mu \nu}(x).
\end{equation}
This implies that if we select $g_{\alpha^\prime \beta^\prime}(x^\prime)$
to be the small-scale metric, it is impossible to obtain an
averaged metric (corresponding to cosmological scales) different from 
the assumed large-scale metric. That is,

\begin{eqnarray}
<g_{\mu\nu}(x)>=\int_{all\,space} 
j^{\alpha^\prime}_{\mu}(x,x^\prime)
j^{\beta^\prime}_{\nu}(x,x^\prime) g_{\alpha^\prime \beta^\prime}(x^\prime)
f(x,x^\prime) d^4x^\prime=\nonumber \\
g_{\mu \nu}(x)
\int_{all\,space} 
f(x,x^\prime) d^4x^\prime=
g_{\mu \nu}(x). 
\end{eqnarray}
This represents a crucial problem for averaging, since, as we have stressed, we 
ideally want to obtain metrics on larger scales as averages
of the small-scale metrics, without assuming the particular form of the
large-scale metrics.  

\item{ Zotov and Stoeger Averaging}

Zotov and Stoeger first presented a simple procedure in which they
average over an
unbound distribution of stars --in a static background--
and of galaxies --in an expanding background 
\cite{Zotov}.
They use an averaging scheme which recalls a three dimensional version of
finding a running mean with the simplest elementary volume: a sphere. Later,
using similar techniques they construct averages over hierarchies of
Swiss-cheese regions in elementary cosmological
cells, which are the smallest volumes which are expanding with the Hubble flow.
Though, they do not provide an averaging procedure which is specifically
defined at each point of cosmological space-time, 
 but rather construct averages over simple
inhomogeneous configurations which they assign to entire regions, one can
argue that over large volumes this procedure approximately gives an average
at each point. In the examples treated, 
the metric (Swiss-cheese) 
of the inhomogeneities is considered spherically symmetric and therefore either
Schwarzschild or FLRW. The attempt is to construct the physical large-scale
background metric from averages over inhomogeneities which may be very large
in amplitude compared to the an provisional background. These averages are
 obtained  by
adding the averages over each inhomogeneity to one another along with the
averages over the provisional ``background'' space in between the
homogeneities. Bi-vectors are not used. We shall come back to discuss this
scheme more fully at the end of the paper, as it can be construed as an
approximate implementation of our averaging procedure in simple non-perturbative
averaging situations.

\item{Buchert and Ehlers averaging of Newtonian Cosmologies}

With a similar approach to that proposed by Noonan, Buchert and Ehlers
\cite{Buchert1,BuchertEhlers} have focused on the averaging problem applied to 
the cosmological expansion of the Universe. They
restrict themselves to the Newtonian approach, i.e. the analog of Friedmann's
equation for the motion of self-gravitating presureless fluid.
They propose that the spatial average
of a tensor field $A$ in the domain $D(t)$ should be:
\begin{equation}
<A>_{D} = \frac{1}{V} \int_{D} d^3x A.
\end{equation}
Since the comoving volume $D(t) = a_D^3(t)$, the fluid elements 
move on the average according to
\begin{equation}
<\theta>_D = \frac{\dot V}{V} = 3 \frac{\dot{a}_D}{a_D}.
\end{equation}
With this in mind, $a_D(t)$ becomes the new scale factor and is shown to
obey an {\it averaged Raychaudhuri equation}.
Note that there are no spatial dependencies in the averaged quantities, only
time is left as a variable. This means that the averaging scheme yields
the same result at all positions in the synchronous gauge.

They also propose a scheme for general relativity which includes the
determinant of the
spatial metric $g = \det g_{ij}$ in the region of interest:
\begin{equation}
<A>_{D} = \frac{1}{V} \int_{D} d^3x \sqrt{g} A
\end{equation}
with $V = \displaystyle{\int}_D d^3x \sqrt{g}$, which resembles Noonan's
definition even more closely. 
This average was then used by Russ et al. \cite{Russ} to compute
the effect of inhomogeneities on the age of a flat Universe. The 
sources of the inhomogeneities were taken from the Zel'dovich approximation to 
second order, so that their results are valid in
an early stage of the evolution of the flat Universe. They conclude
that the effect on the age of the universe is very small on those scales. 

\item{Boersma's Averaging Procedure}

Boersma \cite{Boersma} 
derived, from basic assumptions, a generic linearized
spatial averaging operation for
metric perturbations from FLRW, satisfying the
condition that unperturbed FLRW is a stable fixed point of the averaging
(that is, the averaging operation does not introduce spurious
perturbations in the averaged metric). 
He specified the correspondence 
among points in the real spacetime, the averaged spacetime and the background
spacetime, 
by the introduction of a bi-tensor density in the averaging 
integrals, which fulfils the same function as the bi-vectors in Isaacson's
approach. He 
succeeded in deriving a general form of this bi-tensor density
in terms the background metric and the future-directed unit vectors normal
to the space-like surfaces over which the averages are being performed. With
this formalism, and using Bardeen's \cite{Bardeen} gauge-invariant quantities
Boersma was able 
to resolve the gauge problem in his averaging procedure and
apply it to the constraint equations on spacelike hypersurfaces, which are
closely related to the 
generalized Friedmann equation. 
\end{itemize}

\section{The weak field limit}
\label{WFL}

Using our averager defined in Eq. 
(2) we begin to examine its properties
in a very simple, almost trivial way, by considering its application to 
gravity in the weak field limit. We do this in order to assure ourselves that
our averaging procedure fulfils the simplest intuitive requirements and to
establish some results which we shall need later in applying it to 
perturbed FLRW cosmologies. \\

Let us consider a Minkowskian line element  plus small 
corrections. 
That is,

\begin{equation}
\label{eqdef}
g_{\mu\nu}(x) = \eta_{\mu\nu}(x) + h_{\mu\nu}(x), \qquad |h_{\mu\nu}| \ll 1.
\end{equation}
Applying our averager to this metric we have
\begin{equation}
<g_{\mu\nu}(x)>=\frac{{\displaystyle \int \! d\Omega' \,\sqrt{-g(x + x')} 
\,g_{\mu\nu}(x+x')}}
{{\displaystyle \int \! d\Omega' \,
\sqrt{-g(x + x')}}}.
\end{equation}
Since in the weak field limit (see below)
\begin{equation}
\sqrt{-g(x + x')}=1+\frac12 \,\eta^{\mu \nu}h_{\mu \nu}(x + x'),
\end{equation}
and considering that 
the product of two or more elements of the matrix of perturbations is 
negligible we find, to first (linear) order, 
\begin{equation}
<g_{\mu\nu}(x)>=\eta_{\mu\nu}+\frac{\int \! d\Omega' \,
h_{\mu\nu}(x+x')}
{\int \! d\Omega' } + {\mathcal O}(h^{2}).
\end{equation}
Thus, in general,  averaging a Minkowski space + perturbations 
gives another perturbed Minkowski space -- a result which is obvious, but
reassuring.  It is, of course, possible in some circumstances, that the 
second term of Eq. (9) will be zero, that is, the 
perturbations will average out over larger scales. For instance, 
``overdensities'' may be partially or completely compensated
by ``underdensities''. \\

Furthermore, it is clear that 
in this weak field case -- and in the case of perturbations to FLRW discussed
in Section V -- the operation of averaging {\it will} commute with 
constructing the Einstein field equations, since these are now linear.
Thus, averaging will not introduce any new
terms in the macroscopic field equations themselves, as it does in the exact
case \cite{Ellis,Zotov,Roque}. \\

\label{SecWFL}

Let us now consider a region in space, free of gravitational sources but filled
with gravitational radiation coming from a source situated at infinity. In 
these
circumstances we can apply the weak field limit of Einstein's equations
\begin{equation}
 \nabla h_{\mu \nu} = 0, 
\end{equation}
\begin{equation}
 \frac{\partial h^{\mu}_{\nu}}{\partial x^{\mu}}= \frac{1}{2} \frac{\partial
 h^{\mu}_{\mu}}{\partial x^{\nu}} .
\end{equation}
  
Let us now suppose that we have two scales, a macroscopic and a microscopic
scale, both free of sources. Then on both scales
we will see the same behavior, i.e. on each
scale we can write
\begin{equation}
\label{wflimit}
h_{\mu \nu}^{(i)} = e_{\mu \nu}^{(i)} {\rm e}^{ i k^{(i)}_{\lambda} 
x^{(i) \lambda}} + C.C. 
\end{equation}
for $i=1, 2$, where the $e_{\mu \nu}^{(i)}$ are the polarization tensors in 
each region. 
The microscopic scale is associated with our definition
and the macroscopic
one ought to be compared with the averaged result.

Recalling the definition of the average in the weak field limit, 
\begin{equation}
\label{avh}
 <h_{\mu \nu}(x)> = \frac{\int \! d^{4}x' \,\sqrt{-g(x + x')}\,
  h_{\mu \nu}(x + x')}
{\int \! d^{4}x'\,\sqrt{-g(x + x')} }, 
\end{equation}
where $ g = \det {g_{\mu \nu}} = \det(\eta_{\mu \nu} + h_{\mu \nu})$.
Writing Eq.~(\ref{eqdef}) in matrix notation 
allows us to compute $g$ using


\begin{equation}
\label{logdet}
 \ln \det (1+ A) = tr \ln (1+A).
\end{equation}  
by properly defining $A$ in terms of $h_{\mu \nu}$. Thus 

\begin{eqnarray}
 tr \ln (1+A)& = & \sum_{\alpha} \ln (1+A)_{\alpha \alpha} \nonumber  \\ 
 &  = & \ln(1 + h_{00}) + \ln(-1 + h_{11}) + \ln(-1 + h_{22}) + 
\ln(-1 + h_{33}) \nonumber  \\
& = & \ln(1 + h_{00})(-1 + h_{11})(-1 + h_{22})(-1 + h_{33}) \nonumber  \\ 
 & \approx & \ln(-1 - h_{00}+ h_{11}+ h_{22}+ h_{33}),
\end{eqnarray}
and
\begin{equation}
 -g = 1 + h_{00} - h_{11} - h_{22}- h_{33}, 
\end{equation}
or
\begin{equation}
 \sqrt{-g} \approx 1 + \frac{1}{2}\,\eta^{\mu\nu} h_{\mu\nu}. 
\end{equation}
This result is independent of the choice of the gauge, and
the only hypothesis used is that $h_{\mu \nu}$ is small.
Going back to the definition of the average, and using the property just
demonstrated, we find to first order in $h$
\begin{equation}
<h_{\mu \nu}(x)> = \frac{\int \! d^4 x'\, h_{\mu \nu}(x + x')}{\int \! d^4 x'}.
\end{equation}

Therefore, using Eq.~(\ref{wflimit})
\begin{equation}
<h_{\mu \nu}(x)> = \frac{1}{\Omega'}\biggl[e_{\mu\nu} \int \! d^4 x' \,
{\rm e}^{ i k_{\lambda}\,(x^{\lambda} + x'^{\lambda})} + {e}^{*}_{\mu\nu} 
\int \! d^4 x' \, {\rm e}^{- i k_{\lambda}\,(x^{\lambda} + x'^{\lambda})}
\biggr],
\end{equation}
or
\begin{equation}
<h_{\mu \nu}> = c_{1} \,e_{\mu \nu} \,{\rm e}^{ i k_{\lambda}\,x^{\lambda}} 
+ C.C.
\end{equation}
First of all, note that this averaged quantity  
transforms as a
tensor in the weak field limit. Second the polarization 
tensors can
now be redefined as they appear multiplied by a constant which contains
information about  the microscopic scales of the system. Thirdly,
we have found
that the averaged solution (i.e. the average of the 
solution in one scale)
coincides with the solution on another scale.
This was expected, because we have worked in the linear theory. There 
Einstein equations
are linearized, and therefore, as we have already indicated, the averaging 
procedure does not introduce extra terms in them. This means, of course, that
the averages
are also solutions of the
Einstein weak field equations. \\

We will now extend this analysis to see what happens in the weak field
limit, when considering
a region in which there is a source of gravitational radiation.
In this case, one scale would correspond to that of
the source, while the other could be thought of as the wave zone -- that is,  a
scale in which the length scales are much larger than the size of the source.
When averaging, we should compare the average metric $<{h}_{\mu \nu}>$ with
the one obtained for the wave zone.\\

The solution of the weak field Einstein equations in the presence 
of a source is the metric \cite{Weinberg} 
\begin{equation}
 g_{\mu \nu} = \eta_{\mu \nu} + h_{\mu \nu} 
\end{equation}
with 
\begin{equation}
 h_{\mu \nu}(x) =  4 G \int \! d^{3} x'\frac{S_{\mu \nu}(\vec{x}',
 t-|\vec{x}-\vec{x}'|)}{|\vec{x}-\vec{x}'|}.
\end{equation} 
Combining this
Eq. with Eq. (\ref{avh}) we can write
(to linear order in $h$)
\begin{equation}
\label{averag}
<h_{\mu \nu}(x)> = \frac{4G}{\Omega'} \int \! d^{3} x' d^{4}x'' 
\frac{S_{\mu \nu}
(\vec{x}', 
t + t'' - |\vec{x} + \vec{x}''- \vec{x}'|)}{|\vec{x}+ \vec{x}''-\vec{x}'|}.
\end{equation}

If we now express the energy-momentum tensor as a Fourier integral, we
can analyze each Fourier component separately, and 
integrate (or add those components) afterwards.
This means that we can replace in Eq.~(\ref{averag}), 
$ S_{\mu \nu}(\vec{x},t) = \hat{S}_{\mu \nu}(\vec{x},\omega) 
{\rm e}^{-i \omega t}$, i.e.

\begin{equation}
\label{4}
<h_{\mu \nu}(x)> = \frac{4G}{\Omega'} \int \! d^{3}x'\, d^{4}x'' \, 
\hat{S}_{\mu
\nu}(\vec{x}',\omega) \frac{{\rm e}^{- i \omega\,( 
t + t'' - |\vec{x} + \vec{x}''- \vec{x}'|)}}{|\vec{x}+ \vec{x}''-\vec{x}'|}
+ C.C.
\end{equation}
\mbox{}Hereafter, we drop the complex conjugate of the quantities to
make the expressions look simpler. However, they should be recalled at the end
of the calculations, as all quantities are real.
So far the only approximation made was to consider a
weak field limit.

The scales in the problem are three:
\begin{enumerate}
\item Scale of the source, given by $x'$.
\item Scale of the averaging procedure (or {\em small scale}), given by
$x''$.
\item Scale of the metric ({\em large scale}), 
given mainly by $x$.
\end{enumerate}

This means that scale (1) should be comparable with scale (2), and both much
smaller than scale (3).
Under this assumption we can perform a multipole expansion, which to first
order is 
\begin{equation}
|\vec{x} + \vec{x}''- \vec{x}'| \approx  |\vec{x}- \vec{x}'| + \vec{x}''
\cdot \frac{(\vec{x}- \vec{x}')}{|\vec{x}- \vec{x}'|}.
\end{equation}
Thus, we can integrate over the {\em small scale} and obtain for the spatial
part
\begin{equation}
I = \int \! d^3 x'' \frac{{\rm e}^{ i \omega |\vec{x} + \vec{x}''- \vec{x}'|}}
{|\vec{x}+ \vec{x}''-\vec{x}'|} =
\frac{{\rm e}^{ i \omega |\vec{x} - \vec{x}'|}}{|\vec{x}-\vec{x}'|} \int \! 
d^3 x''
{\rm e}^{ i \omega \vec{x}''
\cdot \frac{(\vec{x}- \vec{x}')}{|\vec{x}- \vec{x}'|}},
\end{equation}
or
\begin{equation}
I = 4 \pi \frac{{\rm e}^{ i \omega |\vec{x} - \vec{x}'|}}
{\omega |\vec{x}-\vec{x}'|}
\left[ \frac{-1}{\omega} R \cos{\omega R} + \frac{1}{w^2}\sin{\omega R}\right],
\end{equation}
where $R$ is the size of the region over which we are averaging. As we
stated before, $R$ should be comparable to the size of the source, and
much smaller than the overall scale of the problem. 
Substituting in Eq.~(\ref{4}), we obtain the averaged metric as
\begin{eqnarray}
\label{6}
& <h_{\mu \nu}(x)> & =  \frac{16 \pi G}{\Omega'} \int \! d t'' {\rm e}^{-i 
\omega \,(t + t'')} \times \nonumber \\
 & & \int \! d^{3} x' \,
\frac{{\rm e}^{ i \omega \,|\vec{x} - \vec{x}'|}}{\omega |\vec{x}-\vec{x}'|}\,
\left[ \frac{-1}{\omega} R \cos{\omega R} + \frac{1}{w^2}\sin{\omega R}\,
\right] 
\,\hat{S}_{\mu \nu}(\vec{x}',\omega). 
\end{eqnarray}
Applying again the multipole expansion for $|\vec{x}'| << |\vec{x}|$, 
finally 
\begin{equation}
\label{7}
<h_{\mu \nu}(x)> = \frac{16 \pi G}{\Omega'} \,{\rm e}^{-i \omega t} 
\,\frac{{\rm e}^{i \omega r}}{r}\,
 \frac{1 - {\rm e}^{-i \omega T}}{i \omega}\,
 \left[\frac{-1}{\omega} R \cos{\omega R} 
+ \frac{1}{w^2}\sin{\omega R}\,\right]\, \int d^3 x'
{\rm e}^{-i \omega \vec{x}' \dot \frac{ \vec{x}}{r}}
\hat{S}_{\mu \nu}(\vec{x}', \omega),
\end{equation}
or,
\begin{equation}
<h_{\mu \nu}(x)> = {\rm e}^{ i k_{\lambda}\, x^{\lambda}}
\,e_{\mu \nu}(\vec{x},\omega) + C. C.
\end{equation}
with 
\begin{equation}
 e_{\mu \nu}(\vec{x},\omega) = \frac{ 4 G}{r}\frac{4 \pi G}{\Omega'} \frac{1 - 
 {\rm e}^{-i \omega T}}{i \omega} \,
\left[\frac{-1}{\omega} R \cos{\omega R} 
+ \frac{1}{w^2}\sin{\omega R}\,\right] \,\int d^3 x'
{\rm e}^{-i \omega \vec{x}' \dot \frac{ \vec{x}}{r}}
\hat{S}_{\mu \nu}(\vec{x}', \omega).
\end{equation}
Here $e_{\mu \nu}$ is the polarization tensor that an observer in the wave 
zone would detect. 

Note the contribution from the
integration over the time $t''$, $(1 - 
 {\rm e}^{-i \omega T})/i \omega $,
with $T$ being the size of the time interval in space-time.
Certainly one expects  the choice of the limit of integration $T$ to be
dependent
on the nature of the problem. Let us suppose that only one frequency $\omega$
is being emitted by a source. Therefore, a suitable choice for $T$ will 
be $\frac{\pi}{\omega}$. That is, $T$ is basically the characteristic period
of the system.
Note as well that we have  only  taken into account 
characteristic scales of the system which are observed from the microscopic 
perspective.

Notice, too, that the averaged metric is, again to first order, a tensor. And
even more, it has the 
same form as the solution of Einstein's equations in the weak field limit
considered in the wave zone.
This was to be expected, because we are still dealing
with the linearized equations, and thus, the averaging is a linear procedure.
That is, averaging the equations is still equivalent to averaging the metric,
just because we are working in the weak-field limit.

\section{The approximate tensorial character of the averaged quantities}
In this section we examine whether or not the quantities defined by our
averaging procedure are generally, or approximately, tensors under
certain conditions. \\

We first analyze the tensorial status of our averaging procedure in the
completely general
case. In order to have $<L_{\mu \nu}>$ as a tensor we would need the 
transformation of $L_{\mu \nu}$ such that 

\begin{equation}
\label{ip}
L^\prime _{\mu \nu} (y + x^\prime) = \frac{\partial y^\alpha}{\partial x^\mu}
\frac{\partial y^\beta}{\partial x^\nu}L_{\alpha \beta} (x+x^\prime)
\end{equation}                                           
under a completely general function $x^\mu=f^\mu(y)$. If Eq.~(\ref{ip}) 
is valid, we can take the derivatives outside the integral of our averaging
definition and find that the averaged quantities behave as tensors too. 
To satisfy Eq.~(\ref{ip}) we need to make a generic
transformation
between $z=x+x^\prime$ and $z^\prime=y+x^\prime$, with a fixed $x^\prime$.
But defining $z^\prime=h(z)$ in this way is equivalent to imposing very
specific constraints
on the original function $f$: 
$$z^\prime=y+x^\prime=f(x)+x^\prime=h(x+x^\prime).$$
This is not possible for every conceivable 
$f$, and thus the averager is not in general
a tensor. \\

Even when we only transform the cosmological coordinate $x$,
so that the averaging procedure yields almost tensorial quantities with
respect to the cosmological coordinates only (the averaging integrations are
only over $x^\prime$, so that the transformations of the cosmological
coordinates $x$ can be taken through the integrals), \cite{Nelson} 
there is an important 
problem. Strictly speaking, we cannot transform the
cosmological coordinates without also transforming the local coordinates and
altering the functional relationship connecting them. Furthermore, 
the integral over the local coordinates $x'$ is not covariant with respect to
those changes. \\

Can we somehow show that the average is {\it approximately} coordinate
covariant in some cases?  
That is indeed true for the 
weak-field and the perturbed-FLRW cases. \\ 

Let $L_{\mu\nu}$ be a tensor. In the weak field 
limit our averager applied to $L_{\mu \nu}$ takes the form
\begin{equation}
<L_{\mu\nu}(x)>=\frac{
\int \! d\Omega' \left( 1 + \frac12 \,\eta^{\alpha \beta} h_{\alpha \beta}(x +
x') \right) L_{\mu\nu}
(x+x')}
{\int \! d\Omega' \left( 1+ \frac12 \,\eta^{\alpha \beta} h_{\alpha \beta}(x +
x') \right)  }.
\end{equation}
To first order in the perturbation this becomes
\begin{eqnarray}
<L_{\mu\nu}(x)> & = & \frac{1}{\Omega'} \int \! d\Omega' \,L_{\mu\nu}
(x+x') +
\frac{1}{\Omega'} \int \! d\Omega' \,\frac12 \,\eta^{\alpha \beta} 
h_{\alpha \beta}(x + x') \, L_{\mu\nu} (x+x') \nonumber \\
& - & \frac{1}{\Omega'^{2}}\int \! d\Omega'\, L_{\mu\nu}(x+x')
\int \! d\Omega''\, \frac12 \,\eta^{\alpha \beta} h_{\alpha \beta}(x + x'').
\end{eqnarray}
where $\Omega' = \int \! d\Omega'$, is the 4-volume and a real number which 
only depends on scale of the averaging. \\

We now employ Isaacson's averaging procedure and compare it with our own.
In the Isaacson's case we have the averaging definition given in 
Eq. (4),
which yields a {\it bona fide} tensor. We also know that the bi-vectors
$j^{\alpha '}_{\mu}(x, x')$ must satisfy 
Eq. (5). In the weak field and
perturbed space-time cases this implies that
\begin{equation}
j^{\alpha '}_{\mu}(x, x') = \delta^{\alpha '}_{\mu} - (1/2)h^{\alpha '}_{\mu}
 (x, x').
\end{equation}
Thus, 
\begin{equation}
\label{isavg}
<T_{\mu \nu}>_I = \int d^4x' T_{\mu \nu}(x')f(x,x') + O(1),
\end{equation}
where $O(1)$ refers to integrals of first order in $h$. Thus, the first term
in this equation is almost a tensor --differs from a tensor by small terms
of order $h$. Then, our averaging procedure gives an
integral like the first term of 
Eq. (\ref{isavg}), except that it
contains a $\sqrt{-g}$ instead of $f(x, x')$ --which makes no difference 
in the tensorial character of the integral in the weak field and 
perturbed field cases. Thus, our procedure also yields an approximate tensor
in both of these cases, relative to the complete coordinate $x + x'$. \\

\section{The FLRW metric + Perturbations}
\label{SecFRW}

Let us now consider 
an FLRW
space-time with perturbations and 
apply the averager to the perturbed metric 
\begin{equation}
\label{pertflrw}
 g_{\mu\nu}(x)=g_{\mu\nu FLRW}(x) + h_{\mu\nu}(x),
\end{equation}
where the FLRW line element in Cartesian coordinate can be written \cite{Bondi} 
\begin{equation}
\label{flrwmet}
ds^2=dt^2- \frac{a^2(t)\,[dx^2+dy^2+dz^3]}{\{1 + \frac{1}{4}k[x^2 + y^2 + z^2]
\}^2}.
\end{equation}
Applying the averaging, we find 
\begin{equation}
\label{metfrw}
<g_{\mu\nu}(x)>=\frac{{\displaystyle \int \! g_{\mu\nu}(x+x') \,
\sqrt{-g(x + x')} d\Omega'}}
{{\displaystyle \int \! \sqrt{-g(x + x')}\, d\Omega'}}.
\end{equation}
Using Eq.~(\ref{logdet}) the 
determinant of the perturbed metric is,
to first order in $h_{\mu\nu}$ 
\begin{equation}
\label{detFRW}
\det g_{\mu\nu}= \det g_{\mu\nu FLRW}\, ( 1 + h_{\mu \nu}g^{\mu\nu}_{FLRW}),
\end{equation}
and replacing this in Eq.~(\ref{metfrw}) we obtain 
\begin{equation}
\label{avFRW}
<g_{\mu\nu}(x)>= \frac{{\displaystyle \int \!\left(g_{\mu\nu FLRW}(x+ x') +
h_{\mu\nu}(x+
x')\right)
\sqrt{-g_{FLRW}(x + x')\left(1 + h_{\mu \nu}g^{\mu\nu}_{FLRW}
\right)} d\Omega'}}
{{\displaystyle \int \!\sqrt{-g_{FLRW}(x + x')\left(1 + h_{\mu \nu}
g^{\mu\nu}_{FLRW}\right)} d\Omega'}}.
\end{equation}

\subsection{What is a perturbation?}

Here we shall define more precisely what we mean by a 
perturbation to an
FLRW metric. We will say that $h_{\mu \nu}$ in Eq.~(\ref{pertflrw}) is a
perturbation if, just as in Eq.~(\ref{eqdef}), 

\begin{equation}
\label{defpert}
|h_{\mu \nu}| \ll 1.
\end{equation}
Thus, if $h_{\mu \nu}$ is a perturbation to FLRW in
one coordinate system, then in any other coordinate system, in which we can
always write the metric as
\begin{equation}
\label{FRWpert}
g'_{\mu \nu}(y) = g'_{\mu\nu FLRW}(y) + h'_{\mu\nu}(y),
\end{equation}
$h'_{\mu \nu}$ is also a perturbation. Though 
$h'_{\mu \nu}$ will naturally
be different from $h_{\mu \nu}$, it will always satisfy Eq.~(\ref{defpert}),
if $h_{\mu \nu}$ itself does so. That is simply because, if we apply a
general coordinate transformation 
\begin{equation}
\label{chFRW}
g'^{\alpha \beta}(y) = \frac{\partial y^{\alpha}}{\partial x^{\mu}}
\frac{\partial y^{\beta}}{\partial x^{\nu}}g^{\mu \nu}(x)
\end{equation} 
to Eq.~(\ref{pertflrw}), we trivially find that the coordinate transformation
Eq.~(\ref{chFRW}) also relates the $h'_{\mu \nu}$ to the $h_{\mu \nu}$, and
that, therefore, the $h'_{\mu \nu}$ will be small, if the $h_{\mu \nu}$ are.
Thus also the determinate of the transformed metric given in
Eq.~(\ref{FRWpert}) will be given by Eq.~(\ref{detFRW}), with the unprimed
metric variables just replaced by the primed (transformed) metric variables.

\subsection{Averaging FLRW Perturbations }

Let us now obtain a general expression for the average of the perturbed
metric. From Eq.~(\ref{avFRW})
\begin{equation}
<g_{\mu\nu}>=<g_{\mu\nu FLRW} + h_{\mu\nu}> = \frac{N}{V},
\end{equation}
with
\begin{eqnarray}
 N & = & \int \! d^4 x' \sqrt{-g_{FLRW}(x + x')} g_{\mu \nu FLRW}(x + x') + 
  \int \! d^4 x'\sqrt{-g_{FLRW}(x + x')} h_{\mu \nu}(x + x') \nonumber \\
& + & \int \! d^4 x' \sqrt{-g_{FLRW}(x + x')} 
\frac{1}{2} g^{\alpha \beta}_{FLRW}(x + x') h_{\alpha \beta}(x + x')
g_{\mu\nu FLRW}(x + x'),
\end{eqnarray}
and
\begin{equation}
V  =  \int \! d^4 x' \sqrt{-g_{FLRW}(x + x')}(1 + \frac{1}{2} 
g^{\alpha \beta}_{FLRW}(x + x') h_{\alpha \beta}(x + x')).
\end{equation}
The denominator can be written as a product, the first term being 
the invariant volume for 
an FLRW
space-time. Expanding the rest of the terms to
first order in $h_{\mu \nu}$ we get
\begin{eqnarray}
\label{av1FRW}
<g_{\mu\nu}> & = & \frac{\int \! d^4 x' \sqrt{-g_{FLRW}(x+x')} g_{\mu\nu FLRW}
(x + x')}{
\int \! d^4 x' \sqrt{-g_{FLRW}(x+x')}} +
\frac{\int \! d^4 x' \sqrt{-g_{FLRW}(x+x')} 
h_{\mu \nu}(x + x')}{\int \! d^4 x' \sqrt{-g_{FLRW}(x+x')}} 
\nonumber \\ 
&+& \frac{\int \! d^4 x'\sqrt{-g_{FLRW}(x+x')} 
\frac{1}{2} g^{\alpha \beta}_{FLRW}(x + x') h_{\alpha \beta}(x + x')
g_{\mu\nu FLRW}(x + x')}{\int \! d^4 x' \sqrt{-g_{FLRW}(x+x')}} \nonumber \\ & - & 
\int \! d^4 x' \sqrt{-g_{FLRW}(x+x')} g_{\mu\nu FLRW}(x + x') \times \nonumber
\\ & \times &
\frac{\int \! d^4 x' \sqrt{-g_{FLRW}(x+x')}\frac{1}{2} g^{\alpha \beta}_{FLRW}(x + x') 
h_{\alpha \beta}(x + x')}{\left(\int \! d^4 x' \sqrt{-g_{FLRW}(x+x')}\right)^{2}}.
\end{eqnarray} \\

We are now in a position to show that, for our averaging procedure, the average
of 
an FLRW metric itself
is, to first order in the local coordinates, 
always an FLRW metric, and further
that the average metric of a perturbed
FLRW metric is also 
an FLRW metric plus perturbations. These are among the
features
any averaging procedure should possess for it to be considered
minimally adequate. \\

First, since the 
length scale represented by the local coordinate $x'$ will
always be much less than that represented by the cosmological coordinates
$x$, we can assume that
the volume of integration, and therefore the domain of $x'$, will
be small with respect to the overall space-time coordinates $x$. Thus,
we can expand the various functions of 
$(x + x')$ as multi-variable
(because of the multiple coordinate variables, $x'$, $y'$, $z'$ and $t'$) 
Taylor series in these local coordinates, and truncate the series at first
order.  We can then write, for instance, 

\begin{equation}
\label{0exp}
g_{\mu \nu FLRW}(x+x') \approx g_{\mu \nu FLRW}(x) + \frac{\partial 
g_{\mu \nu FLRW}}{\partial x^{\lambda}}(x) x'^{\lambda}.
\end{equation}
Analyzing  the first term in Eq.~(\ref{av1FRW}), we find that, in the Cartesian
coordinate form of the metric specified in Eq.~(\ref{flrwmet}),
$<g_{00}^{I}> = 1$. Here and in the discussion which follows, we designate the
four terms in Eq.~(\ref{av1FRW}) by the superscripts
I, II, III, and IV, respectively. Further, using Eq.~(\ref{0exp}) we find also
that
\begin{equation}
<g_{ij}^{I}> = g_{ijFLRW} + \frac{\partial g_{ijFLRW}}{\partial x^{\lambda}}
(x) \frac{\int d^4 x' \sqrt{-g_{FLRW}(x+x')} x'^{\lambda}}
{\int d^4 x' \sqrt{-g_{FLRW}(x+x')}}. 
\end{equation}

These results for the first term of Eq.~(\ref{av1FRW}) enable us to determine
what the application of our averager to an exact FLRW metric yields. If we
carry out the integration indicated in the denominator over the averaging
volume given in the local coordinates $x'^{\lambda}$, we find that this term
vanishes, since the integral is odd in $x'^{\lambda}$. Thus, only even terms
of the Taylor series will yield non-zero contributions to the average. Thus,
to first order in the local coordinates, averaging 
an FLRW metric yields 
an FLRW metric. This result is trivial in the case of averaging a flat FLRW
metric, but is not at first sight
obvious for open and closed FLRW metrics. This
result is somewhat important, because, as Boersma \cite{Boersma} has
emphasized, we do not want the averaging procedure to introduce perturbations
-- we want the average of 
an FLRW metric to be an FLRW metric! \\

If we now go back to 
Eq.~(\ref{av1FRW}), we see that, since the terms II, III and IV will all be
smaller than the pure FLRW term I, we have the further almost
trivial result that
\begin{equation}
<g_{\mu\nu}> = g_{\mu\nu FLRW} + Perturbations.
\end{equation}
Though these are expected and obvious results, they are very reassuring. Our
averaging procedure is physically and mathematically consistent. If it did 
not fulfil these conditions, it would have to be abandoned. \\

Studying terms II, III and IV of Eq.~(\ref{av1FRW}) in more detail, we find
similarly that to first order 
 \begin{equation}
<g_{\mu\nu}^{II}> = \frac{\int \! d^4 x' \sqrt{-g_{FLRW}(x+x')} 
h_{\mu \nu}(x + x')}{\int \! d^4 x' \sqrt{-g_{FLRW}(x+x')}}
\approx h_{\mu \nu}(x) + \frac{\partial h_{\mu \nu}}{\partial x^{\lambda}}(x)
\frac{\int \! d^4 \sqrt{-g_{FLRW}(x+x')} x' x'^{\lambda}}{\int \! d^4
\sqrt{-g_{FLRW}(x+x')}x'},
\end{equation}

and that
\begin{equation}
<g_{\mu \nu}^{III}> = - <g_{\mu \nu}^{IV}> 
\end{equation}
to first order in $h$ and $x'$. \\

Then Eq.~(\ref{av1FRW}) becomes, to first order in the local coordinates, 
\begin{eqnarray}
\label{pertreb}
<g_{\mu \nu}(x)> & = & g_{\mu \nu FLRW}(x) +
\frac{\partial g_{\mu \nu FLRW}}{\partial x^{\lambda}} (x)
\frac{\int d^4 x' \sqrt{-g_{FLRW}(x+x')} x'^{\lambda}}
{\int d^4 x' \sqrt{-g_{FLRW}(x+x')}} + \nonumber \\
  &  & h_{\mu \nu}(x) +
\frac{\partial h_{\mu \nu}}{\partial x^{\lambda}} (x)
\frac{\int d^4 x' \sqrt{-g_{FLRW}(x+x')} x'^{\lambda}}
{\int d^4 x' \sqrt{-g_{FLRW}(x+x')}},
\end{eqnarray}



However, as we just saw above, the integrals in the numerators of the
second and fourth terms of Eq.~(\ref{pertreb}) vanish when taken over the
entire averaging domain, because they are odd in the local coordinates. Thus,
we obtain the somewhat mysterious and apparently trivial result

\begin{equation}
\label{simpres}
<g_{\mu \nu}(x + x')> = g_{\mu \nu FLRW}(x) + h_{\mu \nu} (x),
\end{equation}
to first order in the local coordinates, just what we began with in 
Eq.~(\ref{pertflrw}). It appears that the averaging, to first order in the
local coordinates has no effect -- that the $h_{\mu \nu}$ in
Eq.~(\ref{simpres}) is the same as that in Eq.~(\ref{pertflrw}). This requires
some careful comment. \\

Eq.~(\ref{simpres}) is simply the consequence of terminating the Taylor
series in the local coordinates with the linear term. Any further precision
using this approximation resides in the higher-order terms. In particular,
any further correction to the already small $h_{\mu \nu} (x)$ will be smaller
than it 
already is (that is, of higher order than either $h_{\mu \nu}(x + x')$
itself or the 
value of the local coordinates $x'^{\lambda}$). In general, we should 
expect that $|<h_{\mu \nu}(x + x')>| \leq |h_{\mu \nu}(x)|$, but the
possibility of it being $ < |h_{\mu \nu}(x)|$ can only be explored by either
performing the averaging without approximation, or taking the Taylor series
expansion of the metric to higher orders in $x'^{\lambda}$. \\

It does not make any physical sense to go to higher orders to determine the
distorsions averaging introduces in the FLRW background metric itself (in the
closed and open cases -- if we confine ourselves to averaging over 
spatial hypersurfaces, it is only in the closed and open cases that higher
order distorsions are introduced by averaging),
since it does not vary significantly over local-
coordinate length scales; it is already smooth on all scales up to those of
the cosmological coordinates. It is intuitively clear that averaging over
an FLRW metric should 
give 
an FLRW metric. From one point of view, we might
almost say that the FLRW part of the metric should not be averaged -- it has
in a sense already been averaged, or rather is the result of an average having
been already performed, or assumed -- in particular, an average of the density
used in the
field equations. \\

With respect to the perturbed part of the metric
$h_{\mu \nu}(x + x')$, however, 
there is a good reason to
carry out the averaging to higher precision. In practice, it may, unlike
the FLRW metric itself, vary a great
deal on local length scales. Averaging over those local variations may 
significantly decrease its magnitude on cosmic length scales, even reducing it
to zero. In fact we know that initially there were perturbations on all scales
above that given by photon diffusion (Silk damping). And at our epoch we 
see that there are very large density fluctuations (much larger than
perturbations!) on all local and intermediate scales -- density enhancements
as well as voids. In both cases the 
perturbations on cosmological scales are really the averages over the metric
fluctuations on smaller scales. \\

Thus, when we write the perturbed FLRW metric
as in Eq.~(\ref{pertflrw}) without further specification, we are in a sense
conflating two very different cases, which must be treated separately. The first
is that in which the metric of the universe on {\it all} scales can be
represented
as 
an FLRW metric plus small deviations, which are present on a large
range of scales. In
this case, the expression given in Eq.~(\ref{pertflrw}) is correct on all 
scales. But, at the same time it can be restricted to a given scale of
interest -- for instance, the cosmological scale. In doing that, we would
naturally define the perturbed component $h_{\mu \nu}(x)$ as
\begin{equation}
h_{\mu \nu}(x) = <h_{\mu \nu}(x + x')>,
\end{equation}
where the $h_{\mu \nu}(x + x')$ are small on all length scales and where the
averaging procedure is well-defined as applied to an FLRW mertric plus 
perturbations: As long as the metric 
deviations from the FLRW metric on all scales
are small, the averaging of $h_{\mu \nu}(x + x')$ over a volume of any size
can be performed in a meaningful way, as we have indicated above. That is,
the innumerable small metric fluctuations on local and intermediate scales
can be averaged over to obtain those on cosmic scales. This would be the
situation in the period of the universe up until density perturbations on
one scale or another go nonlinear. It would certainly pertain to the epoch of
recombination when the cosmic microwave background radiation (CMWBR) was last 
scattered, and to some hundreds of millions of years afterwards. Thus, it has
important applications to CMWBR anisotropies. \\

The second situation would be as at present, when the metric of the universe
can be be represented by the perturbed FLRW metric in Eq.~(\ref{pertflrw}) only
on cosmic length scales. On local and intermediate scales Eq.~(\ref{pertflrw})
is far from correct -- the metric on these scales is nowhere near FLRW,
deviations from FLRW being very large. Thus, the interpretation of Eq.~
(\ref{pertflrw}) in this case is that on scales much larger than that of 
any nonlinear density fluctuation, the universe is close to FLRW, if the
$h_{\mu \nu}(x)$ -- the deviations from FLRW on {\it that} scale -- are small.
There will be lower limit on this length scale, below which this will not be
true. Then this $g_{\mu \nu FLRW}(x) + h_{\mu \nu}(x)$ must be understood as 
the average of all
the small-scale  metrics -- some of which will be very different from the
cosmological ``background metric -- over a volume of cosmological length
scale. On small and intermediate length scales we will not be able to define a
background, such that the actual metric is always a perturbation (or small
fluctuation) with respect to it.
How is this $g_{\mu \nu FLRW} + h_{\mu \nu}(x)$
to be calculated from the large amplitude small and intermediate scale
fluctuations? And how should we implement our averaging procedure in this case? 
This is an important question and will be treated briefly in the next section.
\\

In this second case, the averaging procedure we have outlined here is not
at all adequate. But an adequate one must be constructed, if $h_{\mu \nu}(x)$
-- the deviation of the large-scale metric from FLRW -- is to have any meaning
with respect to the metric fluctuations on smaller scales, many of which
are not at all small. What is usually implicitly assumed in using $h_{\mu \nu}
(x)$ in this context is to consider that we have averaged over the 
small and intermediate scale density inhomogeneities to obtain a large-scale
density inhomogeneity which is found to be only marginally different from 
the background FLRW density and is therefore the source of a very small large
scale $h_{\mu \nu}(x)$.. However, this uncritically avoids a number of 
important mathematical and physical issues which really have to be resolved
before such a procedure is validated -- particularly the effective field
equations which are really operative in averaging over large amplitude
inhomogeneities, including the contribution made by gravitational binding
energy to the average, the background metric which is used in the averaging
process, and the adequacy and validity of any averaging process itself 
involving large deviations from either Minkowski or FLRW
\cite{Ellis,Zotov,Zalaletdinov,Roque}. We need to apply our averaging procedure
in these situations and see what it yields. In some simple, idealized cases
an approximate averaging procedure such as that suggested by 
by Zotov and Stoeger \cite{Zotov} may, as we point out below,  be equivalent 
to ours. \\ 

\mbox{}From this brief discussion, we begin to appreciate how different
these two
cases are, from the point of view of averaging -- and the rather different
problems implicit in each of them. As we have seen, it is only the first case,
involving perturbations on all scales which can be dealt with using the
linearized procedures developed so far, or any procedures limited to
perturbations. \\

It is important to note, furthermore, that, as we have already seen, it is
possible that in either of these two cases,
the average of the perturbations or deviations over a large enough scale will
yield zero,
as in some of the weak-field-limit cases. In fact, strictly speaking, if a
given FLRW background model is a genuine background, there should be a large
enough scale over which averaging yields the exact FLRW model itself --
that is, the average over the perturbations on such a large length scale
gives zero. This means that the
``positive'' and ``negative'' perturbations need to balance on the largest
scales, if the universe is really FLRW above a certain length scale. If this
is {\it not} the case, then we must choose the average
FLRW plus perturbations as the real background. It is also clear, of course,
that generally the average of the perturbations will possess much more 
symmetry than the original perturbations themselves. \\

\subsection{Deviations in the Hubble Parameter}

Of course, these issues of averaging will have consequences for the 
observational parameters we measure, for example the Hubble parameter $H$.
It can be easily shown from a perturbation treatment of the field equations
that, for a dust equation of state, the deviation of the Hubble parameter from
its value in a best-fit FLRW model is just given by \cite{Padman}

\begin{equation}
\label{Hubble}
\Delta H = - \frac{1}{3} \dot{\delta},
\end{equation}
where $\delta$ is the density contrast, that is $\displaystyle \delta =
\frac{\rho - \rho_b}
{\rho_b}$, where $\rho$ is the mass-energy density and $\rho_b$ is the 
background (FLRW) mass-energy density. \\

Now, obviously Eq.~(\ref{Hubble}) depends on how $\delta$ is calculated, and
furthermore we want the time-derivative of the 
{\it cosmological $\delta$} -- its
average over the length
scale of the universe. Averages over small or intermediate scales will give
us only the local or intermediate deviations of the Hubble parameter from
its FLRW value. Similar to our discussion of averages over metric
fluctuations in the two general cases in the last subsection, determination
of $\delta$ and therefore $\dot{\delta}$ is relatively easy in the case when
density fluctuations on all scales are perturbations, as long as we have the
required data. Then we can use our procedure to do the averaging. But again, if
we are faced with the second case, where the density inhomogeneities on small
and intermediate scales are large, as at the present time, the correct
averaging procedure of the density fluctuations is unclear. In fact, in
this case, it is not certain that as a consequence of averaging the field
equations at smaller scales over larger scales Eq.~(\ref{Hubble}) holds.
Not only that, but the perturbative treatment from which Eq.~(\ref{Hubble}) is
derived will not be valid in that regime. This issue will be investigated
briefly in the next section, and more
thoroughly in a subsequent paper. Furthermore,
it is obvious again that the scale over which the averaging is done will
determine the result. If we average over intermediate scales -- or determine
$H$ 
using data sampling intermediate scales, scales at which the Hubble flow can
not yet be recovered, instead of cosmological scales -- then our calculation
or measurement of $H$ will be local, not cosmological.

\section{Averaging in Simple Nonperturbative Cases}
When faced with averaging over large amplitude small and intermediate scale
perturbations we can in principle still apply our averaging procedure. However,
in general it is difficult to see how we could practically implement it 
directly. This is true even in simple
idealized cases, where we have, for instance, strong spherically symmetric
inhomogeneities. However,
in such idealized situations, Zotov's and Stoeger's 
two-step procedure \cite{Zotov} provides a useful and implementable
approximation to ours. \\

They first average over each individual spherically symmetric inhomogeneity,
which is represented either as a Schwarzschild metric or a local FLRW metric
plus a surrounding underdense annulus. This average is performed over the
whole vacuole
representing the inhomogeneity, without including any region outside it, 
and is therefore referred to the center point of the inhomogeneity.\\

Now, very large regions of the universe can be ideally considered to be made
up of collections of these spherically symmetric inhomogeneities, or
spherically symmetric clusters of them, separated by regions which are 
either approximately Minkowski -- if the region is not expanding -- or
approximately FLRW, if the region is expanding. These function as a temporary
or intermediate background metric, representing the metric far from the
center of any inhomogeneity or cluster of inhomogeneities. Thus, the picture
is very much like the Swiss-cheese model, except that we envision the
background as temporary or auxilliary -- to enable us to perform the second
step in the averaging. \\

The second averaging is performed over these very large regions --  usually of
cosmological length scale -- consisting of many spherically symmetric
inhomogeneities. It is approximated by:  
\begin{equation}
\label{fulavg}
\bar{g}_{\mu \nu} (x) \cong \frac{1}{V_2}[<g_{\mu \nu}>V_1 N + g_{\mu \nu B}
  (V_2 - V_1 N)],
\end{equation}
where $<g_{\mu \nu}>$ is the average over a single spherically symmetric
homogeneity, $V_1$ is the volume of each homogeneity, and $N$ is the number
of them in the full cosmological averaging volume $V_2$. $g_{\mu \nu B}$ is the
auxilliary background metric. In this above equation we have idealized all the
inhomogeneities as identical in volume and mass. We can easily generalize to
many different sizes. $x$ is the cosmological coordinate at which the
second averaging is centered. \\

Thus, the second averaging can be conceived as done at each value of the
cosmological coordinate $x$. We simply move the volume $V_2$ around the
universe -- to all different values of $x$ and perform the two-step averaging
procedure. It can be easily seen that this is equivalent to what our
averaging procedure involves, precisely in terms of moving the same volume
around the
universe and performing an average over it. Thus it may be considered as a way
of approximating what our procedure would give
in these situations of strong localized inhomogeneities. As Zotov and
Stoeger point out, $\bar{g}_{\mu \nu}(x)$ will generally not be an FLRW 
metric, even if it does not depend on the spatial coordinates -- it will 
superficially look like FLRW, but the scale factor will have a different
dependence on time than the FLRW metric. In general, of course, the result of
the averaging will depend on the spatial cosmological coordinates. \\

Only if
the averaging is over a large enough cosmological volume and the universe is
such that averaging over that volume centered at each point in a spacelike
slice of the universe gives exactly the same result, will it yield a scale
factor independent of the spatial coordinates. If the dependence of the average
scale factor on the spatial coordinates is weak and its time dependence
not too much different from FLRW, then the average metric may be represented
by a perturbed FLRW metric
$g_{\mu \nu FLRW}(x) +  h_{\mu \nu}$, where now $h_{\mu \nu}
$ is a large length-scale perturbation from a FLRW metric which is defined
via the above averaging procedure. However, strictly speaking, we should 
just treat $\bar{g}_{\mu \nu} (x)$ as the average metric over that length
scale, since it will no longer exactly satisfy the Einstein field equations.
\cite{Zotov} It will satisfy field equations which are Einstein's, with an
extra term added, due to the noncommutability of averaging and forming
the Einstein tensor from the metric. \\

Is such averaging approximately tensorial? It seems that it should be, with
respect to the cosmological coordinates. For the prescription for averaging
does not depend on the coordinate system used for the cosmological coordinates
themselves. But do
significant problems arise with regard to the implied averaging over local
coordinates? And is the Zotov-Stoeger averaging over the moving volume really
approximately
equivalent to what our averager would give? We shall investigate these
questions, along with others
in a subsequent paper.

\section{Conclusions}

In this paper we have proposed an intuitively clear averaging procedure for 
general relativity and cosmology, which is an extension to that used in
electromagnetism.
We have shown that it gives approximately unique and tensorial results in
weak field and perturbed FLRW cases, and does not lead to any significant
unacceptable results in these cases. Furthermore, it promises to be
easily applicable in cases where fluctuations on all scales are perturbations,
such as up to the epoch in which density perturbations begin to go nonlinear.\\

Finally, we explore the limits of averaging 
perturbed FLRW universes, indicating that averaging over very large small
and intermediate scale inhomogeneities, in order to recover the average
perturbation on large scales, requires applying our averaging procedure beyond
the simple FLRW plus perturbations case. We show how that can be done
approximately in
simple idealized cases involving spherically symmetric objects and clusters of
objects in an intermediate background using the Zotov and Stoeger approach.


\begin{thebibliography}{99}

\bibitem{Ellis} G. F. R. Ellis in {\it General Relativity and Gravitation} B.
Bertotti
{\it et. al.} eds. (Dordrecht, Reidel, 1984), pp. 215-288.

\bibitem{Isaac} R. Isaacson, Phys. Rev. {\bf 166}, 1263; 1272 (1968).

\bibitem{Noonan1} T. W. Noonan, Gen. Rel. Grav. {\bf 16}, 1103 (1984).

\bibitem{Noonan2} T. W. Noonan, Gen. Rel. Grav. {\bf 17}, 535 (1985).

\bibitem{Zotov} N. V. Zotov and W. R. Stoeger, 
Class. and Quant. Grav. {\bf 9}, 1023 (1992); 
Astrophys. J. {\bf 453}, 574 (1995). 

\bibitem{Boersma} J. P. Boersma, Phys. Rev. D {\bf 57}, 798 (1998). 

\bibitem{Zalaletdinov}  R. M. Zalaletdinov, Gen. Rel. Grav. {\bf 24}, 1015
(1992); {\bf 25}, 673 (1993).

\bibitem{Jackson} J. D. Jackson, {\it Classical Electrodynamics} (J. Wiley
and Sons, 1975), pp. 103-108.

\bibitem{Nelson} A.. H. Nelson, Mon. Not. R. astr. Soc. {\bf 158}, 159 (1972).

\bibitem{EllisMatr} G. F. R. Ellis and D. R. Matravers, Gen. Rel. Grav. 
{\bf 27}, 777 (1995).

\bibitem{EllMaZal} R. Zalaletdinov, R. Tavakol and G. F. R. Ellis, Gen. Rel.
Grav. {\bf 28}. 1251 (1996).

\bibitem{Roque} W. L. Roque, Ph.D. Dissertation, University of Cape Town
(1985).

\bibitem{DeWitt} B. S. DeWitt and R. W. Brehme, Ann. Phys. (N.Y.) {\bf 9}, 220
(1960).

\bibitem{Synge} J. L. Synge, {\it Relativity, The General Theory}, (North
Holland Publ. Co., Amsterdam, 1960) 

\bibitem{Grav} C. W. Misner, K. S. Thorne and J. A. Wheeler, {\it Gravitation}
(W.H. Freeman, 1973)

\bibitem{Buchert1} T. Buchert, in {\it Mapping, Measuring and Modelling the
Universe}, ASP Conference Series, Vol. 94, P. Coles, V. J. Martinez and
M.-J. Pons-Borderia, editors, San Francisco, Astronomical Society of the
Pacific, 1996, pp. 349-356. 

\bibitem{BuchertEhlers} T. Buchert and J. Ehlers, A\&A {\bf 320}, 1 (1997)

\bibitem{Russ} H. Russ, M.H. Soffel, M. Kasai and G. B\"orner, Phys. Rev. D
{\bf 56}, 2044 (1997). 

\bibitem{Bardeen} J. M. Bardeen, Phys. Rev. D {\bf 22}, 1882 (1980)

\bibitem{Weinberg} S. Weinberg, {\it Gravitation and Cosmology} (J.Wiley, 1972)

\bibitem{Bondi} H. Bondi, {\it Cosmology}, Second Edition (Cambridge
University Press, 1968), p. 102. 

\bibitem{Padman} T. Padmanabhan, {\it Structure Formation in the Universe},
Cambridge University Press, 1993, p. 144-145.

\end{thebibliography}
\end{document}